\documentclass[12pt,letterpaper]{article}
\usepackage{setspace,wrapfig}

\usepackage{mathptmx} 
\usepackage{amssymb}
\usepackage{mdframed}
\usepackage{natbib}
\usepackage[english]{babel} % English language hyphenation

\usepackage{microtype} % Better typography

\usepackage{amsmath,amsfonts,amsthm} % Math packages for equations

\usepackage[svgnames]{xcolor} % Enabling colors by their 'svgnames'

\usepackage[hang, small, labelfont=bf, up, textfont=it]{caption} % Custom captions under/above tables and figures

\usepackage{booktabs} % Horizontal rules in tables

\usepackage{lastpage} % Used to determine the number of pages in the document (for "Page X of Total")

\usepackage{graphicx} % Required for adding images

\usepackage{enumitem} % Required for customising lists
\setlist{noitemsep} % Remove spacing between bullet/numbered list elements

\usepackage{sectsty} % Enables custom section titles
\allsectionsfont{\usefont{OT1}{phv}{b}{n}} % Change the font of all section commands (Helvetica)

\usepackage{geometry} % Required for adjusting page dimensions

\usepackage[T1]{fontenc} % Output font encoding for international characters
\usepackage[utf8]{inputenc} % Required for inputting international characters

\usepackage{XCharter} % Use the XCharter font

\usepackage{fancyhdr} % Needed to define custom headers/footers
\fancypagestyle{firstpage}{ % Page style for the first page with the title
	\fancyhf{}
}

\newcommand{\authorstyle}[1]{{\large\usefont{OT1}{phv}{b}{n}\color{MidnightBlue}#1}} % Authors style (Helvetica)

\newcommand{\institution}[1]{{\small\usefont{OT1}{phv}{m}{sl}\color{Black}#1}} % Institutions style (Helvetica)

\usepackage{titling} % Allows custom title configuration

\newcommand{\HorRule}{\color{DarkGoldenrod}\rule{\linewidth}{1pt}} % Defines the gold horizontal rule around the title

\pretitle{
	\vspace{-50pt} % Move the entire title section up
	\HorRule\vspace{-30pt} % Horizontal rule before the title
	\fontsize{30}{36}\usefont{OT1}{phv}{b}{n}\selectfont % Helvetica
	\color{MidnightBlue} % Text colour for the title and author(s)
\begin{flushleft}
}

\posttitle{\par\end{flushleft}\vskip 10pt} % Whitespace under the title

\preauthor{\begin{flushleft}} % Anything that will appear before \author is printed

\postauthor{ % Anything that will appear after \author is printed
\end{flushleft}
	\vspace{10pt} % Space before the rule
	\par\HorRule % Horizontal rule after the title
	\vspace{10pt} % Space after the title section
}

\usepackage{lettrine} % Package to accentuate the first letter of the text (lettrine)
\usepackage{fix-cm}	% Fixes the height of the lettrine

\newcommand{\initial}[1]{ % Defines the command and style for the lettrine
	\lettrine[lines=3,findent=4pt,nindent=0pt]{% Lettrine takes up 3 lines, the text to the right of it is indented 4pt and further indenting of lines 2+ is stopped
		\color{DarkGoldenrod}% Lettrine colour
		{#1}% The letter
	}{}%
}

\usepackage{xstring} % Required for string manipulation

\newcommand{\lettrineabstract}[1]{
	\StrLeft{#1}{1}[\firstletter] % Capture the first letter of the abstract for the lettrine
	\initial{\firstletter}\textbf{\StrGobbleLeft{#1}{1}} % Print the abstract with the first letter as a lettrine and the rest in bold
}

\usepackage{natbib}
\title{X-ray Studies of Planetary Systems: A 2020 Decadal Survey White Paper}
\author{\authorstyle{
\noindent
Jaesub Hong\textsuperscript{1*}, 
Suzanne Romaine\textsuperscript{2}, 
Larry Nittler\textsuperscript{3},
Martin Elvis\textsuperscript{2}, 
Ian Crawford\textsuperscript{4},
Graziella Branduardi-Raymont\textsuperscript{5},
Lucy Rim\textsuperscript{6},
Scott Wolk\textsuperscript{2}
}
\newline\newline % Space before institutions
\textsuperscript{1}\institution{Harvard University},
\textsuperscript{2}\institution{Smithsonian Astrophysical Observatory},
\textsuperscript{3}\institution{Carnegie Institution of Washington},
\textsuperscript{4}\institution{Birbeck College, University of London, UK},
\textsuperscript{5}\institution{University College London, UK},
\textsuperscript{6}\institution{NASA Goddard Space Flight Center},
\textsuperscript{}\institution{*Corresponding author: jhong@cfa.harvard.edu} 
}

\date{A White Paper submitted to the {\em The Planetary Science and Astrobiology Decadal Survey 2023-2032} Committee, National Academies of Sciences, Engineering and Medicine\\ July 15, 2020}

\begin{document}

\maketitle
\thispagestyle{firstpage} % Apply the page style for the first page (no headers and footers)
\newpage

%- short intro to advances in X-ray technology
%  [ref] 2016 Hong et al EPS paper
  
%- PICASSO: VV questions to text
%- CubeX: what high resolution imager can do as an example

\lettrineabstract{Whether it is fluorescence emission from asteroids and moons, solar wind charge exchange from comets, exospheric escape from Mars, pion reactions on Venus, sprite lighting on Saturn, or the Io plasma torus in the Jovian magnetosphere, the Solar System is surprisingly rich and diverse in X-ray emitting objects.} 
The compositions of diverse planetary bodies are of fundamental interest to planetary science, providing clues to the formation and evolutionary history of the target bodies and the solar system as a whole. X-ray fluorescence (XRF) lines, triggered either by solar X-rays or energetic ions, are intrinsic to atomic energy levels and carry an unambiguous signature of the elemental composition of the emitting bodies. Except for H, He and Li, all the atomic elements are identifiable by their K, L or M shell XRF lines in the 0.1 to 15 keV band. Light elements including life-essential C, N, and O fluoresce in soft X-rays, whereas heavier elements (like the important rock-forming elements Mg, Al, Si, S, Ca, Ti, Cr, and Fe) emit higher energy (harder) X-rays. XRF is a powerful diagnostic tool to understand the chemical and mineralogical composition of planetary bodies. All remote-sensing XRF spectrometers used so far on planetary orbiters have been collimated instruments, with limited achievable spatial resolution, and many have used archaic X-ray detectors with poor energy resolution. Focusing X-ray optics provide true spectroscopic imaging and are used widely in astrophysics missions, but until now their mass and volume have been too large for resource-limited in-situ planetary missions. Recent advances in X-ray instrumentation such as the Micro-Pore Optics used on the {\it BepiColombo} X-ray instrument \citep{Fraser10}, Miniature X-ray Optics \citep{Hong16} and highly radiation tolerant CMOS X-ray sensors \citep[e.g.,][]{Kenter12} enable compact, yet powerful, truly focusing X-ray Imaging Spectrometers. NASA PSD has funded some of the recent developments in XRF technology through programs like PICASSO. Such instruments will enable compositional measurements of planetary bodies with much better spatial resolution and thus open a large new discovery space in planetary science,  greatly enhancing our understanding of the nature and origin of diverse planetary bodies. Here, we discuss many examples of the power of XRF to address key science questions across the solar system. {\bf We urge NASA to continue to provide robust support to the development of novel X-ray instrumentation and to encourage the consideration of such instruments for diverse planetary missions in the coming decade.}

%}

\setcounter{page}{1}

%moved Moon section here to introduce cubex concept as an example of power of imaging telescopes htat we can then refer to in later sections

\section{The Moon}
\begin{figure}
%\centering % Using \begin{figure*} makes the figure take up the entire width of the page
\includegraphics[width=0.55\textwidth]{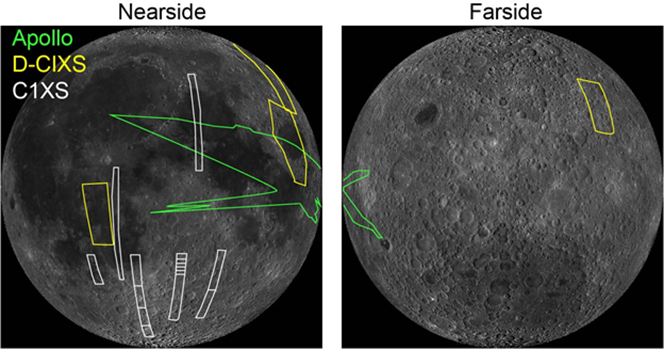}
\caption{Lunar regions investigated by previous XRF mapping missions outlined on LRO \citep{Vondrak10} Camera lunar basemap in orthographic projection. Spatial resolution of these data is larger than ~30 km. Almost all of the sparse previous coverage is on the nearside.
}
\label{f:lunarcoverage}
%\vspace{-0.0in}
\end{figure}

The Moon was the first target for remote-sensing XRF measurements, on Apollo 15 and 16 \citep{Adler72} and several subsequent missions have carried X-ray instruments for measuring the surface elemental composition of the Moon, with mixed success \citep[e.g.,][]{Grande09}.  Despite the proximity to Earth, the spatial coverage of all lunar XRF experiments to date is quite sparse and of relatively poor spatial resolution (>~30 km) as illustrated in Figure~\ref{f:lunarcoverage}, due in part to the use of collimated instruments. Global elemental abundance maps (e.g., K, Th, Fe, Ti), produced from  remote-sensing $\gamma$-ray, neutron, and reflectance spectroscopy data, have been extremely useful for understanding the Moon's history. However, XRF spectroscopy extends measurements to geochemically highly important elements like Mg and Al, and the promise of higher-resolution measurements enabled by focusing X-ray optics will enable new alternative approaches with higher success rates. High resolution mapping at a few km scale on the Moon would help to address multiple important  science questions: What are the major surface features and modification processes on each of the inner planets? What governed the accretion, supply of water, chemistry, and internal differentiation of the inner planets and the evolution of their atmospheres, and what roles did bombardment by large projectiles play? What are the distribution and timescale of volcanism on the inner planets? 

As an example of a next-generation planetary XRF application, {\it CubeX} is a MiXO-telescope-based SmallSat lunar mission, whose concept was studied with support from NASA’s Planetary Science Deep Space SmallSat Studies (PSDS3) program \citep{Hong18, Kashyap20}.  {\it CubeX} is a SmallSat with a compact, highly radiation-tolerant, focusing X-ray telescope that identifies and spatially maps lunar crust and mantle materials excavated by impact craters, and also serves as a pathfinder for autonomous precision deep-space navigation using X-ray pulsars.

Instead of staying at a low lunar orbit ($\lesssim$100 km altitude), a MiXO telescope like {\it CubeX} can stay at a “frozen” high-altitude ($\gtrsim$6000 km) orbit and monitor a handful of targeted patches with several hours per orbit spent pointing at each. With the ability to distinguish elemental composition with better than 3-km resolution within a $\sim$110-km FoV, such a mission can address many questions of lunar evolution, including the identification of outcrops of excavated mantle and lower crust material to probe magma ocean differentiation, studies of the vertical composition of near and farside crust, and exploration of volcanic domes and flows to understand the geochemical evolution of the mantle and crust.
Compared to the {\it Chandrayaan}/D-C1XS instrument \citep{Grande09}, the reduced X-ray flux seen by the MiXO telescope due to the larger distance and the smaller FoV is largely made up by the larger effective area, lower background and longer exposure for a given FoV footprint. A typical solar flare of any class lasts a few minutes to hours. A MiXO telescope in a high altitude orbit can monitor the same patch of the surface during the entire duration of a typical flare whereas D-C1XS cannot monitor the same patch for more than $\sim$20 sec with a FoV “footprint” of $\sim$25 km. By repeatedly viewing the same targets on each orbit, {\it CubeX} can build up statistics for each, resulting in high-resolution elemental maps of scientifically important targets. Simulations indicate that with 500 ksec total integration, easily obtainable in a one-year mission, {\it CubeX} can determine Mg/Si, Al/Si, Fe/Mg and other elemental ratios for six sites with relative precision ranging from 5\% on a 100 km scale to 30\% on a 3 km scale at C1 solar states, sufficient to distinguish diverse lunar compositions and address the science questions outlined above. A similar instrument flown on a full-scale lunar orbiter could do much, much more.

\section{Asteroids}

Primitive bodies, be they comets, asteroids, Trojans, or Kuiper Belt objects, provide windows to study our early Solar System. For instance, many asteroids including the parent bodies of chondritic meteorites were not subjected to planetary differentiation and melting, and thus retain a record of the primordial nebula composition, including primordial organic matter. Recent discoveries of compositionally diverse main belt asteroids indicate substantial mixing through dynamical processes such as early planetary migration \citep{DeMeo14}. Discoveries of several main belt “asteroids” exhibiting cometary behavior (e.g., 133P/Elst-Pizarro) imply that the main belt contains a large dormant population of volatile-rich asteroids. Some of the key questions that may be answerable through XRF compositional measurements include: How do spectral variations among asteroids relate to compositional differences and how do these relate to disk and/or planetary evolution and differentiation? How diverse were the components from which asteroids were assembled? What relationships exist between asteroid and meteorite classes?  How did differentiation vary on bodies with large proportions of metal or ices?  Were there radial or planetesimal-size limits on differentiation? A comparative study of the composition of diverse asteroids vs.~larger planets and moons through XRF spectroscopy can help answer these questions, revealing the formation and evolutionary histories of these bodies and the Solar System. 

There are only two prior X-ray observations of asteroids, 433 Eros by {\it NEAR Shoemaker} \citep{Trombka00}, and 25143 Itokawa by Hayabusa \citep{Okada06}. In addition, {\it OSIRIS-REx}, currently exploring asteroid 101955 Bennu, includes an X-ray spectrometer (REXIS) but no data have yet been reported. The prior X-ray measurements were made with mechanically collimated detectors – gas proportional counters on NEAR and X-ray CCDs on Hayabusa – both with ~3.5–5 degree spatial resolution of their field of view (FoV). Even with its poor spectral resolution ($\sim$830 eV FWHM at 5.9 keV), {\it NEAR} measured major abundance ratios such as Mg/Si, Al/Si and Ca/Si that supported a long-suspected connection between ordinary chondrite meteorites and S-type asteroids and the hypothesis that space weathering modifies the apparent composition of asteroids as seen by optical and infrared techniques \citep{Nittler01}. Definitive evidence of space weathering alteration has been later identified in the returned Itokawa samples \citep{Noguichi11}. These results addressed a decades old conundrum of the mismatch between the spectra of meteorites and common asteroids. 

\begin{figure}[htb]
%\centering % Using \begin{figure*} makes the figure take up the entire width of the page
\includegraphics[width=0.545\textwidth]{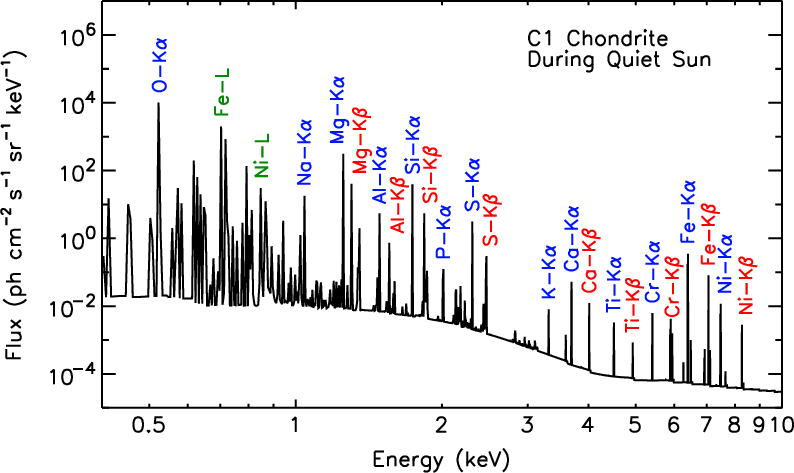}
\includegraphics[width=0.455\textwidth]{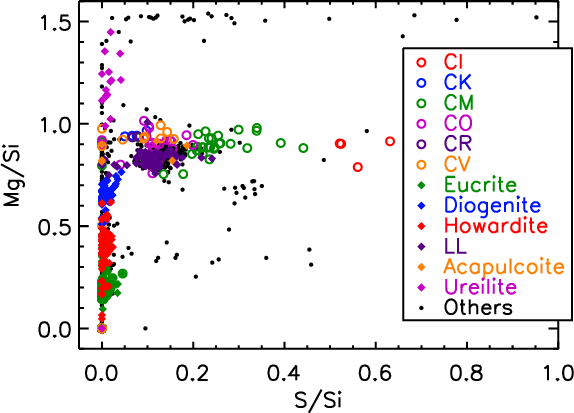}
\caption{(Left) Simulated XRF spectrum of an asteroid of CI chondrite composition at 1 AU during a quiet sun state illustrating wide range of measureable elements. (Right) Abundance ratios (Mg/Si vs.~S/Si) as an identifier of a wide range of meteorite specimen types and geological processes  \citep{Nittler04}.
}
\label{f:abundance}
%\vspace{-0.0in}
\end{figure}

Imaging spectroscopy via a {\it CubeX}-style instrument would enable the measurement of spatial variation of both the absolute and relative abundances (e.g., Figure~\ref{f:abundance}) of geochemically important elements with high resolution. Such measurements could reveal compositional differences across asteroids due to impact excavation of distinct material, variable presence of exogenic material or history of space weathering, and/or differentiation, all of which would provide new insights into the origin and evolution of small bodies. Moreover, such instruments could prove to be invaluable tools for selecting sampling sites for small-body sample return missions.

\section{Comets}
Like asteroids, comets are relics of the earliest stages of solar system formation, but accreted farther from the Sun and include a better-preserved record of primordial ices and organics. Rosetta provided chemical measurements of dust and gas in the coma of comet 67P/Churyumov-Gerasimenko and at the site of the Philae lander \citep[e.g.,][]{Taylor17}, but not direct elemental measurements of the nucleus. Moreover, it has been known since the surprise {\it ROSAT} discovery of the comet C/Hyakutake1996 B2 \citep{Lisse96}  that many comets are bright in X-rays due to solar wind charge exchange in the coma tracing both incoming solar wind plasma and outflowing neutral gas. About two dozen X-ray bright comets have now been discovered \citep[e.g.,][]{Wolk09}. The observed cometary X-ray emission relative to the optical is ~100 times higher than the Sun. Focusing X-ray telescopes near the target would see 10$^2$–10$^4$ times higher X-ray flux than the near-Earth telescopes and will capture the dynamics of the coma from a 1000 km scale all the way down to sub-km size of the nucleus during a close approach.  

{\bf Comet nuclei}, inaccessible from near-Earth telescopes, are primary targets for future in-situ X-ray study both by dedicated orbiting missions and sample-return missions aimed at returning  pristine icy samples with potential prebiotic organics. Primary science questions that could be addressed by XRF  include: How variable are the comet compositions, and how heterogeneous are individual comets? What kinds of surface evolution, radiation chemistry, and surface-atmosphere interactions occur on distant icy primitive bodies? How is the surface composition of comets modified by thermal radiation and impact processes? In-situ X-ray spectra will isolate the elemental composition of the comet nucleus from the coma, which is optically thin in X-rays and thus can be spatially separated from the nucleus by imaging. The surface distribution of elemental composition will provide valuable input and context for sample collection and provide constraints on ice:dust ratio. Time resolved imaging could capture dynamic features in the nucleus such as outgassing. 

\section{Mercury}
\begin{figure}[h]
%\centering % Using \begin{figure*} makes the figure take up the entire width of the page
\includegraphics[width=0.4\textwidth]{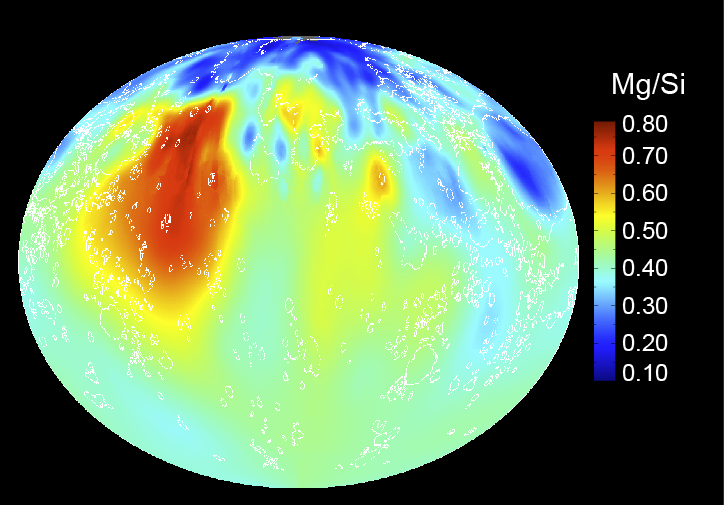}
\caption{Map of Mg/Si across Mercury’s surface with outlines of smooth plains in white from {\it MESSENGER} XRS data \citep{Nittler18b}.
}
\label{f:mercury}
%\vspace{-0.0in}
\end{figure}

{\it MESSENGER} orbited Mercury for over 4 years (2011-2015). Despite the poor spectral resolution with no imaging capability and highly variable spatial resolution, the {\it MESSENGER}  X-ray spectrometer (XRS) produced maps of major-element abundance ratios Mg/Si, Al/Si, S/Si, Ca/Si and Fe/Si across the surface and constrained the abundances of the minor elements Ti, Mn, and Cr \citep{Nittler18b}. These data revealed that Mercury’s surface is surprisingly rich in S and Mg and very low in Fe, indicating the planet formed under much more chemically reducing conditions than did the other terrestrial planets. The abundance maps show remarkable variability (Figure~\ref{f:mercury}) indicating that Mercury’s thin mantle is chemically heterogeneous, providing clues to the planet’s early geological history. The possible presence of wt\% levels of C at the surface would have important implications for a possible magma ocean phase.  {\it BepiColombo}, launched in October 2018 for Mercury, takes advantage of both collimator and focusing optics with two X-ray telescopes onboard \citep{Fraser10}. MIXS-T on {\it BepiColombo}, enabled by micro-pore optics (MPO) will be the first in-situ planetary true imaging X-ray telescope, and will produce high resolution maps around interesting regions such as pyroclastic volcanic deposits and “hollows” (enigmatic depressions likely caused by volatile loss) during solar flares. High-resolution X-ray mapping in future missions could address many questions on the origin and evolution of the planet's surface and could provide regional context to landed Mercury science missions \citep{ByrneWP}.

\section{The Galilean Satellites}
The Galilean Satellites are also excellent targets for X-ray surface composition study, as {\it Chandra} discovered X-ray emission from them in 1999 and 2000 \citep{Elsner02,Nulsen20,Branduardi17}. These satellites are very faint when observed from Earth orbit and the X-ray detections of Io and Europa, while statistically significant, were based on just 10 photons each, a tribute to the value of high angular resolution in X-ray observations. The  detected X-rays had energies that clustered  between 0.5 and 0.7 keV, suggestive of O-K XRF emission. The X-rays from Europa are best explained by fluorescence induced by bombardment of energetic H, O, and S ions on the icy surface  \citep[e.g.,][]{Cooper01}. Ion sputtering is considered to be a major agent for neutral H$_2$O release and the total net erosion of the surface \citep{Plainaki10}. With 1000 times higher flux than near-Earth telescopes, X-ray observations of the Galilean Satellites with focusing X-ray instruments would address these key questions: How is surface material modified exogenically versus being pristine or relatively unmodified? How do exogenic processes control the distribution of chemical species of satellite surfaces? How, and to what extent, have volatiles been lost from Io? What is the temporal and spatial variability of the density and composition of Io's atmosphere, how is it controlled, and how is it affected by changes in volcanic activity? 

The surface compositions of the icy moons are expected to be rich in interesting chemical species \citep[e.g.,][]{,McKinnon03}, and their surfaces \citep{Carlson05} and associated atmospheres \citep{McGrath04} display significant spatial variability. Observed time variations in UV absorption indicate that the relative surface composition for Europa, Ganymede, and Callisto \citep[e.g.,][]{Domingue98a}    vary with time in the irradiated upper layers, probably mostly due to changes in magnetospheric activity but potentially also due to transient emergence of material from internal sources. X-ray imaging spectroscopy enables measurements of major and trace species (e.g., C, N, O, Na, Na, Mg, S, Cl, K, P, Fe) needed to distinguish between models for surface and interior geology, atmospheric evolution, and the presence of subsurface oceans strongly suggested by Galileo magnetometer results, addressing these science questions: Are organics present on the surface of Europa, and if so, what is their provenance? Has material from a subsurface Europa ocean been transported to the surface, and if so, how? Such studies provide clues to an ultimate question: Does (or did) life exist below the surface of Europa or Enceladus? For example, XRF imaging could identify regions on Europa where upwelling and refreezing of ocean water enhances Na, Cl, and Mg abundances, especially if sub-1\% level elemental determination can be attained.

\section{Summary}

Miniaturization of X-ray optics, advances in X-ray sensors, and  
rapid developments of low cost CubeSat/SmallSat spacecrafts and their components
will open a new era of X-ray imaging spectroscopy 
of solar system objects. High resolution mapping of surface elemental abundance 
of diverse planetary bodies will play a critical role in revealing 
the complex history of the formation and evolution of diverse planetary bodies.


\begin{thebibliography}{}
\small
\bibitem[Adler et al.(1972)]{Adler72}
Adler, I., et al., 1972, Science, 177, 256. 

\bibitem[Binzel et al.(2010)]{Binzel10}
Binzel, R. P., et al., 2010, Nature, 463, 331. 

\bibitem[Boehnhardt et al.(2017)]{Boehnhardt17}
Boehnhardt, H., et al., 2017, Philosophical Transactions of the Royal Society A, 375, no. 2097, id.20160248. 

\bibitem[Branduardi-Raymont(2017)]{Branduardi17}
Branduardi-Raymont, G., 2017, Astronomische Nachrichten, 338, 188. 

\bibitem[Byrne et al.(2020)]{ByrneWP}
Byrne, P., et al., 2020, Mercury Landed Science White Paper submitted to Decadal Survey

\bibitem[Carlson et al.(2005)]{Carlson05} 
Carlson, R. W., et al., 2005, ICARUS, 177, 461. 

\bibitem[Cooper et al.(2001)]{Cooper01} 
Cooper, J. F., et al., 2001, ICARUS, 149, 133. 

\bibitem[Crawford et al.(2009)]{Crawford09}
Crawford, A. I., et al., 2009, Planetary and Space Science, 57, 725. 

\bibitem[DeMeo \& Carry(2014)]{DeMeo14}
DeMeo, F. E., and Carry, B., 2014, Science, 505, 629. 

\bibitem[Domingue \& Lane(1998)]{Domingue98a} 
Domingue, D. L., and Lane, A. L., 1998, Geophysical Research Letters, 25, 4421. 

\bibitem[Elsner et al.(2002)]{Elsner02} 
Elsner, R. F., et al., 2002, The Astrophysical Journal, 572, 1077. 

\bibitem[Fraser et al.(2010)]{Fraser10}
Fraser, G. W., et al., 2010, Planetary and Space Science, 58, 79. 

\bibitem[Grande et al.(2009)]{Grande09}
Grande, M., et al., 2009, Planetary and Space Science, 57, 717. 

\bibitem[Hong et al.(2018)]{Hong18} 
Hong, J., et al., 49th Lunar and Planetary Science Conference, Woodlands, Texas, 2018. 

\bibitem[Hong et al.(2016)]{Hong16} 
Hong, J., Romaine, S., and the MiXO Team, 2016, Earth, Planets and Space, 68, 35. 

\bibitem[Hsieh \& Jewitt(2006)]{Hsieh06}
Hsieh, H. H., and Jewitt D., 2006, Science, 312, 561. 

\bibitem[Hsieh et al.(2013)]{Hsieh13}
Hsieh, H. H., et al., 2013, The Astrophysical Journal Letter, 771, 1. 

\bibitem[Johansson et al.(1995)]{Johansson95} 
Johansson, S. A. E., Campbell, J. L., and Malmqvist, K. G., "Particle Induced X-Ray Emission," in Wiley, New York, 1995. 

\bibitem[Kashyap et al.(2020)]{Kashyap20} 
Kashyap, V. L., et al., 2020, Applied Optics, 59, 5560. 

\bibitem[Kenter et al.(2012)]{Kenter12} 
Kenter, A. T., Kraft, R., and Murray, S. S., 2012, SPIE, 8453, 84530G. 

\bibitem[Lantz, Binzel, \& DeMeo(2018)]{Lantz18}
Lantz, C., Binzel, R. P., and DeMeo, F. E., 2018, Icarus, 302, 10. 

\bibitem[Lisse et al.(1996)]{Lisse96} 
Lisse, C. M., et al., 1996, Science, 274, 205. 

\bibitem[McGrath et al.(2004)]{McGrath04} 
McGrath, M. A., et al., 2004, "Satellite Atmospheres , in Jupiter, the Planet, Satellites and Magnetosphere," Cambridge, Cambridge University Press, p. 457.

\bibitem[McKinnon \& Zolensky(2003)]{McKinnon03} 
McKinnon, W. B., and Zolensky, M. E., 2003, Astrobiology, 3, 879. 

\bibitem[Nittler et al.(2018)]{Nittler18b} 
Nittler, L. R., Chabot, N. L., Grove T. L., and Peplowski, P. N., 2018, "2. The chemical composition of Mercury," in Mecury: The view after MESSENGER, Cambridge, Cambridge University Press. 

\bibitem[Nittler et al.(2001)]{Nittler01} 
%Nittler, L. R., Starr, R. D., Lim, L., McCoy, T., and Burbine, T. H., 2001, Meteorites \& Planetary Science, 36, 1673.
Nittler, L. R., et al., 2001, Meteorites \& Planetary Science, 36, 1673.

\bibitem[Nittler et al.(2004)]{Nittler04} 
Nittler, L. R., et al., 2004, Antartic Meteorite Research, 17, 231. 

\bibitem[Noguichi et al.(2011)]{Noguichi11}
Noguichi, T., et al., 2011, Science, 333, 1121. 

\bibitem[Nulsen et al.(2020)]{Nulsen20} 
Nulsen, S., et al., 2020,  The Astrophysical Journal, 895, 79.

\bibitem[Okada et al.(2006)]{Okada06}
Okada, T., et al., 2006, Science, 312, 1338. 

\bibitem[Paranicas et al.(1999)]{Paranicas99} 
Paranicas, C., et al., 1999, Journal of Geophysics Research, 104, 17459. 

\bibitem[Paranicas et al.(2002)]{Paranicas02} 
Paranicas, C., et al., 2002, Geophyiscal Research Letters, 29, 18. 

\bibitem[Plainaki et al.(2010)]{Plainaki10} 
Plainaki, C., et al., 2010, ICARUS, 210, 385. 

\bibitem[Saur et al.(1998)]{Saur98} 
Saur, J., Strobel, D. F., and Neubauer, F. M., 1998, Journal of Geophysical Research, 103, 19947. 

\bibitem[Schmitt et al.(1991)]{Schmitt91}
 Schmitt, J. H. M. M., et al., 1991, Nature, 349, 583. 

\bibitem[Taylor et al.(2017)]{Taylor17} 
Taylor, M. G. G. T., Altobelli, N., Buratti, B. J., and Choukroun, M., 2017, Philosophical Transactions of the Royal Society A, 375, no. 2097, id.20160262. 

\bibitem[Tompson et al.(2019)]{Tompson19} 
Tompson, M. S., et al., Icarus, 319, 499.

\bibitem[Trombka et al.(2000)]{Trombka00} 
Trombka, J. I., et al, 2000, Science, 289, 2101. 

\bibitem[Vondrak et al.(2010)]{Vondrak10} 
Vondrak, R., Keller, J., Chin, G., and Garvin, J., 2010, Space Science Reviews, 150(1-4), 7.

\bibitem[Wolk et al.(2009)]{Wolk09} 
Wolk, S. J., et al, 2009, The Astrophysical Journal, 694, 1293. 

%G. Branduardi-Raymont et al., "Latest results on Jovian disk X-rays from XMM-Newton," Planetary and Space Science, vol. 55, pp. 1126-1134, 2007. 
%G. Branduardi-Raymont et al., "A study of Jupiter's aurorae with XMM-Newton," Astronomy and Astrophysics, vol. 463, p. 761, 2007. 

\end{thebibliography}
\end{document}